\documentclass{article}
\usepackage{amssymb,amsmath,epsfig,epstopdf,color,graphicx,subfigure}
\usepackage{epstopdf,array}
\newcolumntype{L}{>{\centering\arraybackslash}m{3cm}}

\usepackage{arxiv}

\usepackage[utf8]{inputenc} % allow utf-8 input
\usepackage[T1]{fontenc}    % use 8-bit T1 fonts
\usepackage{hyperref}       % hyperlinks
\usepackage{url}            % simple URL typesetting
\usepackage{booktabs}       % professional-quality tables
\usepackage{amsfonts}       % blackboard math symbols
\usepackage{nicefrac}       % compact symbols for 1/2, etc.
\usepackage{microtype}      % microtypography

\usepackage{amsmath,amssymb,amsfonts}
\usepackage{algorithmic}
\usepackage{graphicx}
\usepackage{textcomp}
\usepackage{comment,color}
\usepackage{enumitem}
\usepackage{epsfig}
\usepackage{caption}
\usepackage{verbatim,bm}
\usepackage{epstopdf,array,mathtools,url,slashbox,subfigure}

\newcolumntype{L}{>{\centering\arraybackslash}m{3cm}}
\DeclarePairedDelimiter\norm{\lVert}{\rVert}%
\makeatletter
\let\oldabs\abs
\def\abs{\@ifstar{\oldabs}{\oldabs*}}
\let\oldnorm\norm
\def\norm{\@ifstar{\oldnorm}{\oldnorm*}}
\makeatother
\title{Seismic traveltime simulation for variable velocity models using physics-informed Fourier neural operator}

\author{
  Chao Song\\
 Department of Geophysics, College of Geo-\\
 Exploration Science and Technology, \\
 Jilin University
    \And
   Tianshuo Zhao\\
 Department of Geophysics, College of Geo-\\
 Exploration Science and Technology, \\
 Jilin University
    \And 
  Umair bin Waheed\\
  Department of Geosciences\\
  King Fahd University of Petroleum and Minerals
    \And  
  Cai Liu\\
 Department of Geophysics, College of Geo-\\
 Exploration Science and Technology, \\
 Jilin University
    \And  
   You Tian\\
 Department of Geophysics, College of Geo-\\
 Exploration Science and Technology, \\
 Jilin University
    \And  
}

\begin{document}
\maketitle

\begin{abstract}

Seismic traveltime is critical information conveyed by seismic waves, widely utilized in various geophysical applications. Conventionally, the simulation of seismic traveltime involves solving the eikonal equation. However, the efficiency of traditional numerical solvers is hindered, as they are typically capable of simulating seismic traveltime for only a single source at a time. Recently, deep learning tools, particularly physics-informed neural networks (PINNs), have proven effective in simulating seismic traveltimes for multiple sources. Nonetheless, PINNs face challenges such as limited generalization capabilities across different models and difficulties in training convergence. To address these issues, we have developed a method for simulating multi-source seismic traveltimes in variable velocity models using a deep-learning technique, known as the physics-informed Fourier neural operator (PIFNO). The PIFNO-based method for seismic traveltime generation takes both velocity and background traveltime as inputs, generating the perturbation traveltime as the output. This method incorporates a factorized eikonal equation as the loss function and relies solely on physical laws, eliminating the need for labeled training data. We demonstrate that our proposed method is not only effective in calculating seismic traveltimes for velocity models used during training but also shows promising prediction capabilities for test velocity models. We validate these features using velocity models from the OpenFWI dataset.

\end{abstract}

% keywords can be removed

\section{Introduction}

Seismic traveltime simulation has long been an important topic in geophysics across different scales. Traveltime fields have wide applications in numerous geophysical processes including, but not limited to, normal moveout (NMO) correction \cite{Yilmaz2001Seismic}, seismic tomography \cite{li2013first}, microseismic event estimation \cite{izzatullah2022laplace}. The foundation of seismic traveltime simulation is the eikonal equation, derived from the ray-theoretical approximation of the scalar wave equation \cite{Yilmaz2001Seismic}. It is a non-linear first-order partial differential equation (PDE). Primarily, two numerical methods are most commonly used for solving this PDE: the fast marching method (FMM) \cite{sethian1996fast} and the fast sweeping method (FSM) \cite{zhao2005fast}. The accuracy of these numerical solvers for the eikonal equation heavily depends on the discretization schemes used for spatial and temporal dimensions. Finer meshing schemes enhance the accuracy of seismic traveltime simulations. However, they also substantially increase computational demands, particularly when dealing with multiple sources and various velocity models.

In recent years, deep learning has increasingly demonstrated its efficacy and success within the geophysical community, emerging as a promising tool for solving PDEs \cite{mousavi2022deep,wu2023sensing}. Among various approaches, physics-informed neural networks (PINNs) and neural operators stand out as leading techniques for addressing both forward and inverse problems associated with PDEs \cite{raissi2019physics,kovachki2021neural}. PINNs use spatial and temporal coordinates as input and target function as output. More importantly, instead of relying solely on data mapping, PINNs incorporate physical laws within their loss functions. This innovative approach has led to successful applications in seismic traveltime, as well as seismic wavefield modeling and inversion \cite{alkhalifah2021wavefield,song2021solving,bin2021pinneik,waheed2021pinntomo,rasht2022physics,song2022wavefield}. 
 Despite these advancements, PINNs are also facing several limitations. Typically, they require an extensive number of training epochs, resulting in high computational demands. Additionally, their applicability is often restricted to specific physical models, and they exhibit limited generalization capabilities when applied to varying models. This constraint poses a significant challenge in expanding their utility across different geophysical fields.
 
To address the limitations in model generalization and training difficulty inherent in PINNs, we develop a DL-based method utilizing a physics-informed Fourier neural operator (PIFNO) for predicting seismic traveltimes across various velocity models with multiple sources. Our innovative PIFNO solver integrates the velocity models and corresponding background traveltimes (indicative of source locations) as inputs, and outputs the perturbation traveltime. We employ the factorized eikonal equation as our loss function. When the training is completed, the PIFNO is capable of directly producing seismic traveltime solutions that correspond to given input velocity models and background traveltimes. Significantly, this trained PIFNO model is not only useful for the velocity models included in the training but can also be promptly applied to new test velocity models, predicting seismic traveltimes effectively. We validate the performance and generalization capabilities of PIFNO by testing it on various velocity model families from the OpenFWI dataset \cite{deng2021openfwi}, demonstrating its robustness and adaptability in seismic traveltime prediction.

\section{Theory}

\subsection{The eikonal equation and its factorized form}

Simulating the seismic traveltime relies on solving the eikonal equation, which is a first-order, hyperbolic form of PDE expressed as \cite{julian1977three,cerveny2001seismic}:
\begin{eqnarray}
\left | \nabla T(\mathbf{x}) \right |^{2}=\frac{1}{v^{2}(\mathbf{x})},\; \; \forall \mathbf{x}\in \Omega ,
\label{eqn:eq1}
\end{eqnarray}
where $T(\mathbf{x})$ is the seismic traveltime in the domain of $\Omega$, and $v(\mathbf{x})$ is the velocity model that controls $T(\mathbf{x})$. $\mathbf{x}=\left \{ x ,z\right \}$ specifies the Euclidean coordinates for both the horizontal and vertical positions. $\nabla=\frac{\partial}{\partial x}+\frac{\partial}{\partial z}$ denotes the operator of the first-order spatial derivative. In the source location $\mathbf{x_{s}}$, there exits the source singularity problem, expressed as $T(\mathbf{x_{s}})=0$. To mitigate this problem,  we decompose the traveltime $T(\mathbf{x})$ into two terms: a background traveltime term $T_{0}(\mathbf{x})$ corresponding to a constant velocity and a perturbation traveltime term $\tau$, given by:
\begin{eqnarray}
T(\mathbf{x})=T_{0}(\mathbf{x})+\tau(\mathbf{x}).
\label{eqn:eq2}
\end{eqnarray}
The background traveltime $T_{0}(\mathbf{x})$ can be easily calculated by:
\begin{eqnarray}
T_{0}(\mathbf{x})=\frac{\left | \mathbf{x}-\mathbf{x_{s}} \right |}{v_{0}},
\label{eqn:eq3}
\end{eqnarray}
where $\left | \mathbf{x}-\mathbf{x_{s}} \right |$ denotes the seismic wave travelling distance from the point $\mathbf{x}$ to the source point $\mathbf{x_{s}}$ and $v_{0}$ denotes the velocity value at the source location. The background traveltime can reflect the locations of the sources, which can be easily calculated analytically. If we plug equation \ref{eqn:eq2} into equation \ref{eqn:eq1}, we can get a factorized form of the eikonal equation, given by:
\begin{eqnarray}
\left | \nabla T_{0}(\mathbf{x}) + \nabla \tau(\mathbf{x}) \right |^{2}=\frac{1}{v^{2}(\mathbf{x})}.
\label{eqn:eq4}
\end{eqnarray}
According to the central finite difference scheme, the first-order spatial derivative of $T$, $\frac{\partial T}{\partial x}$ and $\frac{\partial T}{\partial z}$, can be expressed as:
\begin{eqnarray}
\frac{\partial T}{\partial x}\approx \frac{T(iz,ix+1)-T(iz,ix-1)}{2dx}=\begin{bmatrix}
0 & 0 &0 \\ 
 -\frac{1}{2dx}& 0 &\frac{1}{2dx} \\ 
0 & 0 &0 
\end{bmatrix}*\begin{bmatrix}
T(iz-1,ix-1) & T(iz-1,ix) &T(iz-1,ix+1) \\ 
T(iz,ix-1)& T(iz,ix) &T(iz,ix+1) \\ 
T(iz+1,ix-1) & T(iz+1,ix) &T(iz+1,ix+1) 
\end{bmatrix},
\label{eqn:eq5}
\end{eqnarray}

\begin{eqnarray}
\frac{\partial T}{\partial z}\approx \frac{T(iz+1,ix)-T(iz-1,ix)}{2dz}=\begin{bmatrix}
0 &  -\frac{1}{2dz}&0 \\ 
 0& 0 &0\\ 
0 & \frac{1}{2dz}  &0 
\end{bmatrix}*\begin{bmatrix}
T(iz-1,ix-1) & T(iz-1,ix) &T(iz-1,ix+1) \\ 
T(iz,ix-1)& T(iz,ix) &T(iz,ix+1) \\ 
T(iz+1,ix-1) & T(iz+1,ix) &T(iz+1,ix+1) 
\end{bmatrix},
\label{eqn:eq6}
\end{eqnarray}

where $dx$ and $dz$ represent the spatial sampling in the horizontal and vertical directions. This can be also applied to $\nabla \tau$. Equations \ref{eqn:eq5} and \ref{eqn:eq6} can be easily implemented in the PyTorch platform by defining $\begin{bmatrix}
0 & 0 &0 \\ 
 -\frac{1}{2dx}& 0 &\frac{1}{2dx} \\ 
0 & 0 &0 
\end{bmatrix}$ and $\begin{bmatrix}
0 &  -\frac{1}{2dz}&0 \\ 
 0& 0 &0\\ 
0 & \frac{1}{2dz}  &0 
\end{bmatrix}$ as finite-difference kernels convolving with $T_{0}$ and $\tau$.

\subsection{Physics-informed Fourier neural operator}

FNO is an advanced neural operator approach capable of achieving function mapping within infinite-dimensional spaces \cite{li2020fourier}. It has gained wide popularity in solving variable PDEs for geophysical problems \cite{yang2021seismic,song2022high,wei2022small}. Utilizing the factorized form of the eikonal equation (as shown in equation \ref{eqn:eq4}), we find that the velocity model ($v$) and background traveltime ($T_{0}$) together can uniquely determine the perturbation traveltime ($\tau$). Accordingly, our objective in using FNO is to construct a mapping from $\left [ v;\: T_{0} \right ]$ to $\tau$. This mapping process begins by applying a linear operator $P$ to elevate the input $\left [ v;\: T_{0} \right ]$ into a higher-dimensional space, denoted as: $p_{0}=P(\left [ v;\; T_{0} \right ])$. Within the FNO framework, $p_{0}$ is designated as the initial state. This state then evolves through a series of iterative neural network layers, following the sequence $p_{0}\rightarrow p_{1}...\rightarrow p_{j}\rightarrow...\rightarrow p_{T}$. Subsequently, $p_{T}$ is fed into a fully-connected neural network (FCNN) composed of multiple hidden layers. At the conclusion of this process, another linear operator is used to project the FCNN’s output back to the dimension of our target output, $\tau$. Since both the input $\left [ v;\: T_{0} \right ]$ and output $\tau$ are governed by the physics laws, namely the factorized eikonal equation (referenced as equation \ref{eqn:eq4}), we train the network using a physics-constrained loss function, expressed as: 
\begin{eqnarray}
L:=\left \| \nabla T_{0}(\mathbf{x}) + \nabla \tau(\mathbf{x})-\frac{1}{v^{2}(\mathbf{x})} \right \|_{2}^{2}.
\label{eqn:eq7}
\end{eqnarray}

\begin{figure}
\begin{center}
\includegraphics[width=1.0\textwidth]{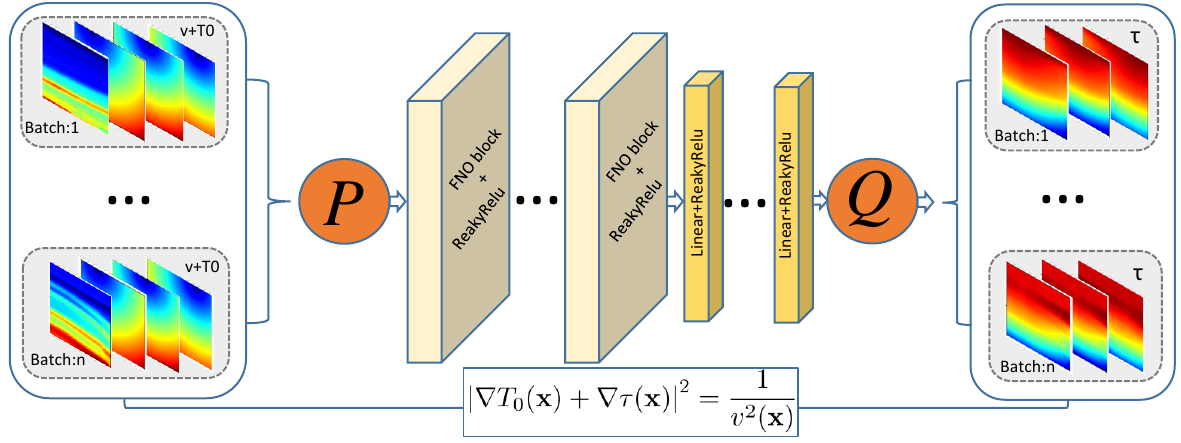} 
\caption{An illustration of the PIFNO architecture for simulating seismic traveltimes. The model uses batches of velocity and background traveltime data as input, processes it through multiple Physics-Informed Fourier Neural Operator blocks with LeakyReLU activation functions, and outputs the perturbation traveltime. The training process is governed by the factorized eikonal equation, showcasing the model's adherence to physical laws.}
\label{fig:PIFNO}
\end{center}   
\end{figure}

This function not only ensures compliance with the underlying physical principles but also guides the network towards accurately predicting the seismic traveltime perturbations: $\tau$. At the source location, the perturbation traveltime $\tau$ should indeed be zero to reflect the immediate emission of seismic waves at the point of origin. To account for this, a hard-constraint condition is imposed within the model: $\tau(\mathbf{x_{s}})=0$. The architecture of the PIFNO is detailed in Fig.\ref{fig:PIFNO}.

The key step in FNO is the iterative evolvement from state $p_{j}$ to $p_{j+1}$. Mathematically, this step is defined by:
\begin{eqnarray}
p_{j+1}(\mathbf{x})=\sigma \left ( Wp_{j}(\mathbf{x})+(K(\phi)p_{j})(\mathbf{x}) \right ),
\label{eqn:eq8}
\end{eqnarray}
with 
\begin{eqnarray}
(K(\phi)p_{j})(\mathbf{x})=\int \kappa _{\phi }(x,y)p_{j}(y)dy,
\label{eqn:eq9}
\end{eqnarray}
where $\sigma()$ is the activation function, and we use LeakyRelu activation function in our experiments; $\kappa _{\phi }$ represents a neural network parameterized by learnable parameters $\phi$; and $W$ is also a learnable matrix used for linear transformation.

Assuming the defined $\kappa _{\phi }$ satisfying this condition, $\kappa _{\phi }(x,y)=\kappa _{\phi }(x-y)$, the integral operation in equation \ref{eqn:eq9} becomes a convolution given by:
\begin{eqnarray}
(K(\phi)p_{j})(\mathbf{x})=\int \kappa _{\phi }(x-y)p_{j}(y)dy.
\label{eqn:eq10}
\end{eqnarray}

\begin{figure}
\begin{center}
\includegraphics[width=0.5\textwidth]{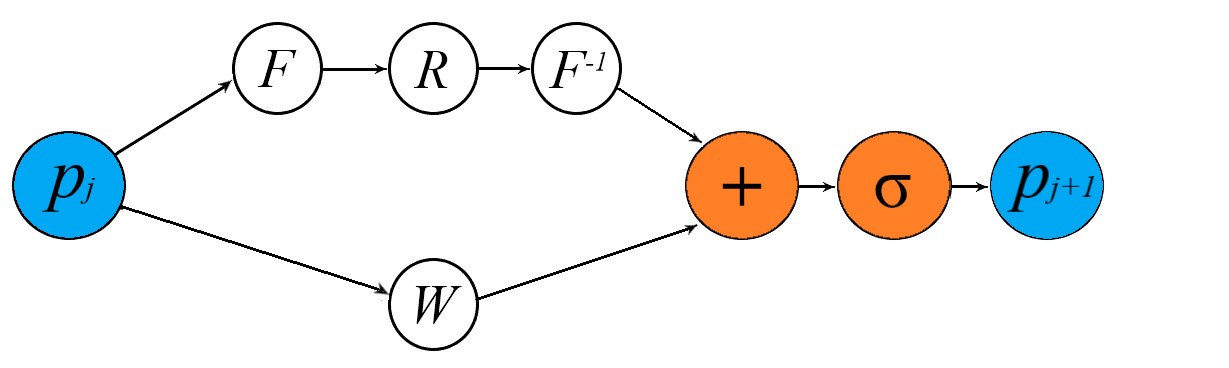} 
\caption{One FNO block with $p_{j}$ and $p_{j+1}$ representing the input and output state.}
\label{fig:FNO_block}
\end{center}   
\end{figure}

In Fourier transformed space, the convolution of two functions becomes the product of their respective Fourier transforms. Thus, equation \ref{eqn:eq9} can be written as:
\begin{eqnarray}
(K(\phi)p_{j})(\mathbf{x})=F^{-1}\left [ F[\kappa_{\phi}(x)] F[p_{j}(x)]\right ],
\label{eqn:eq11}
\end{eqnarray}
where $F$ and $F^{-1}$ denote the forward and inverse Fourier transform. Due to the product implementation in Fourier space, this neural operator is named as FNO \cite{li2020fourier}. In equation \ref{eqn:eq10}, instead of transforming $\kappa_{\phi}(x)$ into the Fourier space, we can directly define complex-valued tensor $R_{\phi}$ to replace $F[\kappa_{\phi}(x)]$. This $R_{\phi}$ can be de parameterized as a low-pass filter to suppress
undesired high-frequency components in $F[p_{j}(x)]$. By doing so, the number of trainable parameters is greatly reduced and the convergence of training is accelerated. We define the process of equation \ref{eqn:eq8} as one FNO block, which is shown in Fig \ref{fig:FNO_block}. In this paper, we use five FNO blocks and five hidden layers in our developed PIFNO architecture for all the experiments.

\subsection{Training setup}

\begin{figure}
\begin{center}
\includegraphics[width=1.0\textwidth]{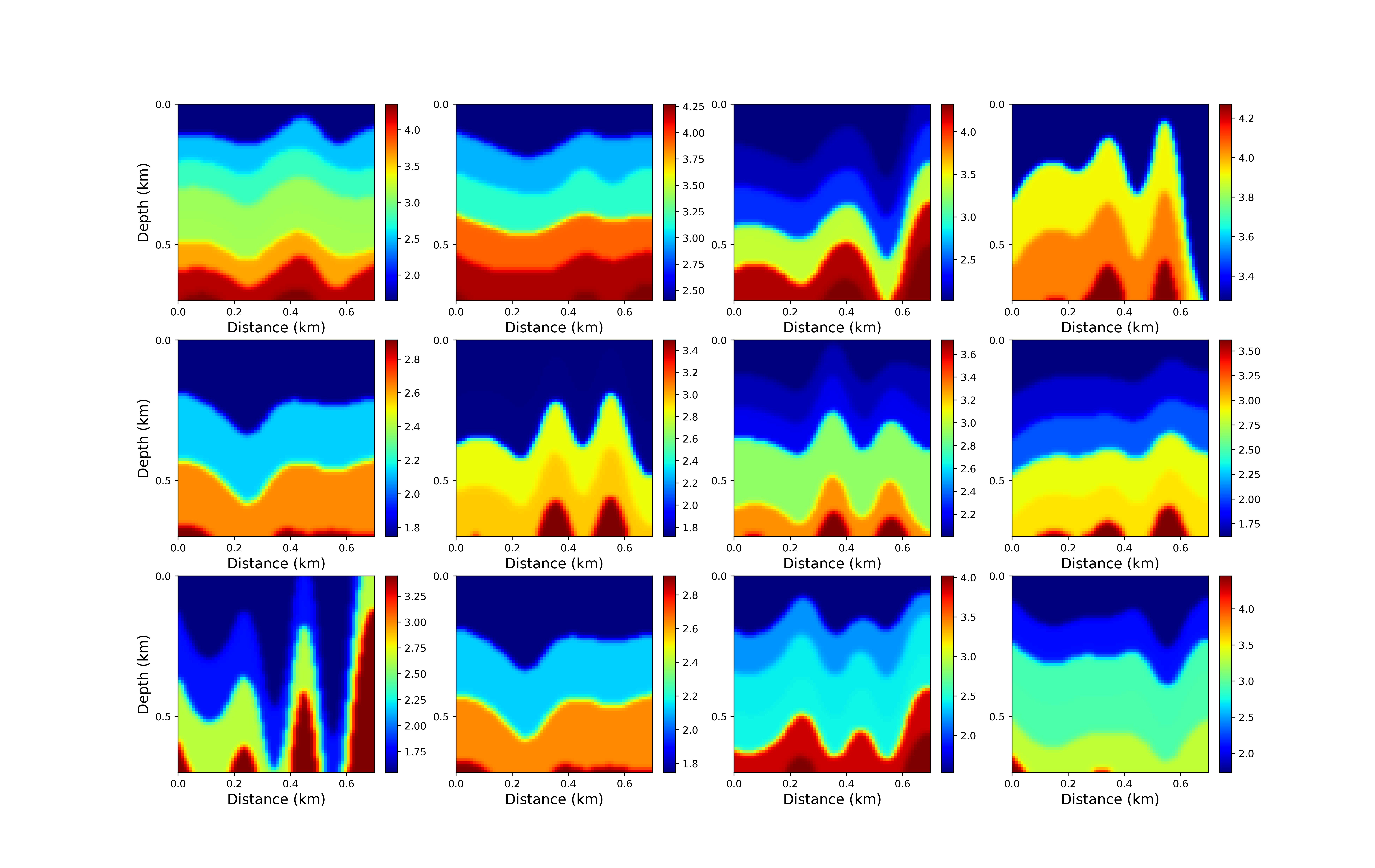} 
\caption{Representative velocity models from the CurveVel-A family.}
\label{fig:CurveVel}
\end{center}   
\end{figure}

The training of the developed PIFNO is conducted through an unsupervised learning approach. This means that there's no requirement for labeled training data, significantly simplifying the data preparation process. The primary goal of the PIFNO is to establish a mapping between the inputs $\left [ v;\: T_{0} \right ]$ and output $\tau$, as shown in Fig. \ref{fig:PIFNO}. Once this mapping is successfully learned, PIFNO can predict the seismic traveltime for any given velocity model with sources located at arbitrary positions. To evaluate the efficacy of our proposed method, we conducted tests using the CurveVel-A velocity family from the OpenFWI dataset \cite{deng2021openfwi}. Representative models from this dataset are shown in in Fig. \ref{fig:CurveVel}. These models display characteristic curved layer structures within the velocity profiles. Each model has dimensions of $70\times70$, with a spatial sampling interval of 10 m in both the vertical and horizontal directions. The capability to predict traveltime in such varied and structurally complex models will reveal the potential of PIFNO in handling diverse geophysical scenarios. 

We use 50 velocity models from the the CurveVel-A velocity family to train the PIFNO. For each velocity model, we generate background traveltimes $T_{0}$ for 10 sources and these sources are uniformly distributed on the surface of the models. We arrange these velocity models and background traveltimes into a five-dimensional (5D) tensor as training data. Each dimension of this input 5D tensor is arranged as follows:
\begin{itemize}
  \item [1)] 
  First Dimension: Batch number (total velocity models) = 50.      
  \item [2)]
  Second Dimension: Source number (total traveltimes) = 10.
  \item [3)]
  Third Dimension: Grid points in the vertical axis = 70. 
  \item [4)]
  Fourth Dimension: Grid points in the horizontal axis = 70. 
  \item [5)]
  Fifth Dimension: Channel number. This includes the background traveltime for each velocity and the velocity model itself, repeated per source number, resulting in 20 channels.
\end{itemize}

The developed PIFNO method streamlines the data processing by compressing the 5D input tensor into a 4D output. This compression is achieved by reducing the channel number from 20 to just one, while retaining the sizes of the output tensor's first four dimensions to match those of the input's. 

\begin{figure}
\begin{center}
\includegraphics[width=0.5\textwidth]{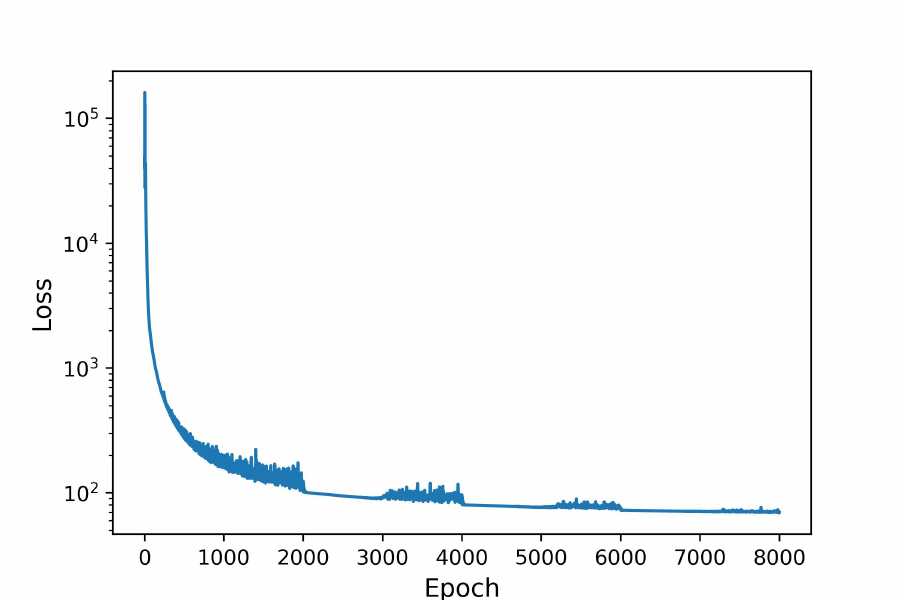} 
\caption{Training loss curve.}
\label{fig:Loss_curvelet}
\end{center}   
\end{figure}

For training the PIFNO model, we employed the Adam optimizer with 8,000 epochs. The training commenced with an initial learning rate of 0.0025, which was halved at every 2,000-epoch interval. We utilized a full-batch training approach for this process. The training loss curve is depicted in Fig. \ref{fig:Loss_curvelet}. Upon examination, we noted the achievement of satisfactory training convergence, indicating the effectiveness of our training methodology and parameter settings.

\section{Results}

\begin{figure}
\begin{center}
\includegraphics[width=1.0\textwidth]{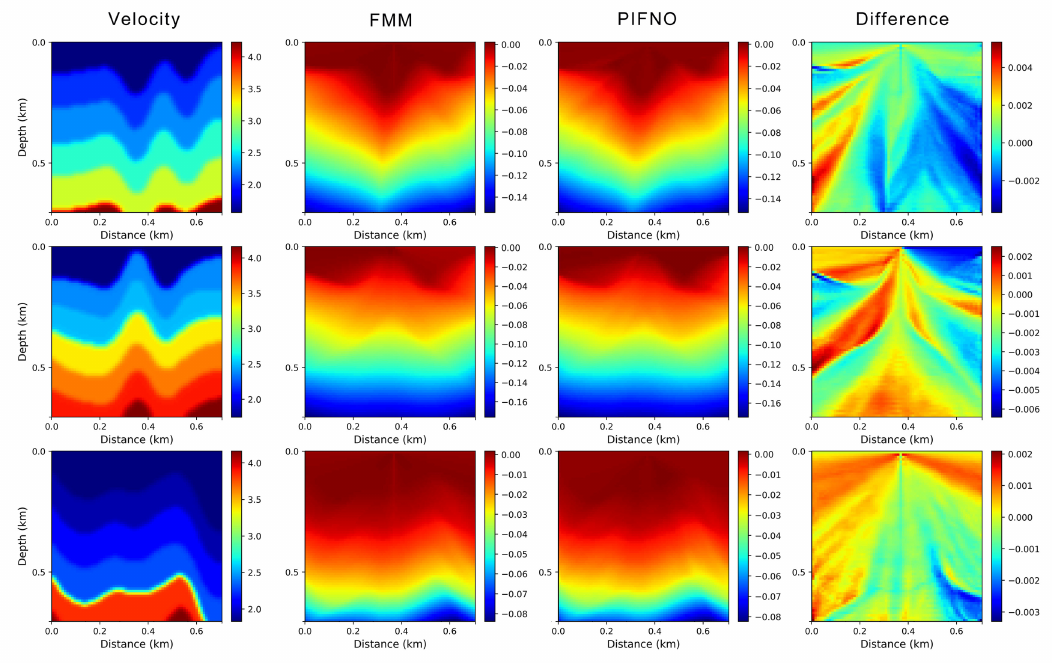} 
\caption{Training data from the CurveVel-A family: velocity models (first column), numerical $\tau$ from FMM (second column), predicted $\tau$ from PIFNO (third column), and $\tau$ difference (fourth column) for a source in the middle.}
\label{fig:curvelet50_train_tau}
\end{center}   
\end{figure}

\begin{figure}
\begin{center}
\includegraphics[width=1.0\textwidth]{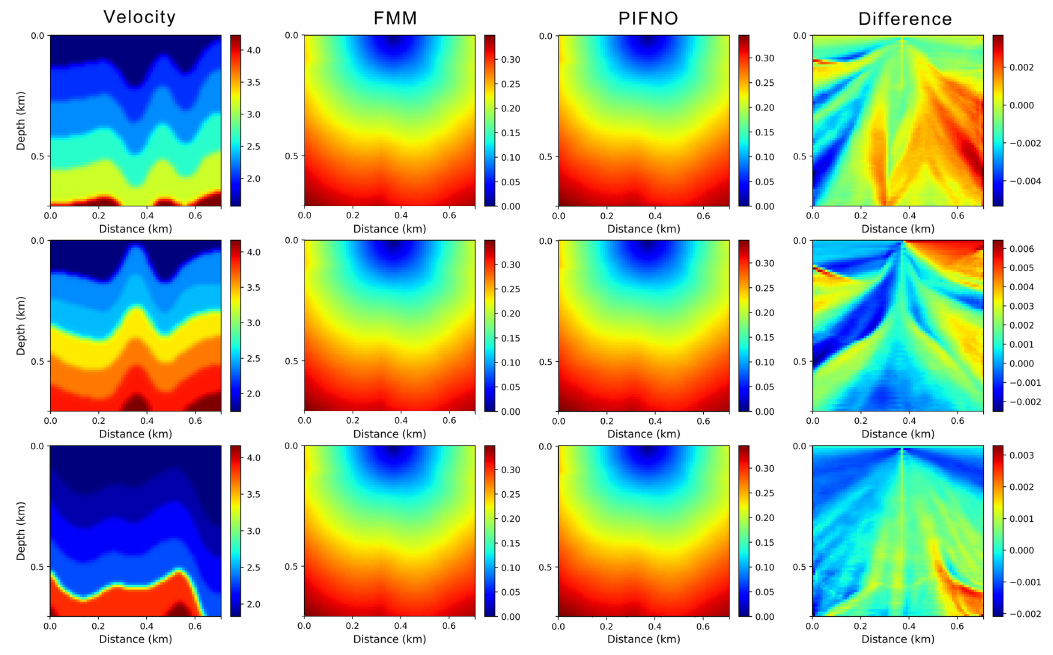} 
\caption{Training data from the CurveVel-A family: velocity models (first column), numerical traveltime from FMM (second column), predicted traveltime from PIFNO (third column), and traveltime difference (fourth column) for a source in the middle .}
\label{fig:curvelet50_train_T}
\end{center}   
\end{figure}

In this section, we evaluate the performance of the PIFNO. Fig. \ref{fig:curvelet50_train_tau} illustrates the experiment setup; the first column presents three different velocity models used in training. The second column depicts the FMM-calculated arrival times ($\tau$) as numerical reference solutions, derived using the scikit-fmm toolkit \cite{skfmm}. In contrast, the third column of Fig. \ref{fig:curvelet50_train_tau} displays $\tau $ as predicted by PIFNO. The similarity between $\tau$ shown in these two columns is satisfactory, with ignorable differences, further validated by the very small discrepancy values in the fourth column of Fig. \ref{fig:curvelet50_train_T}. 

\begin{figure}
\begin{center}
\includegraphics[width=1.0\textwidth]{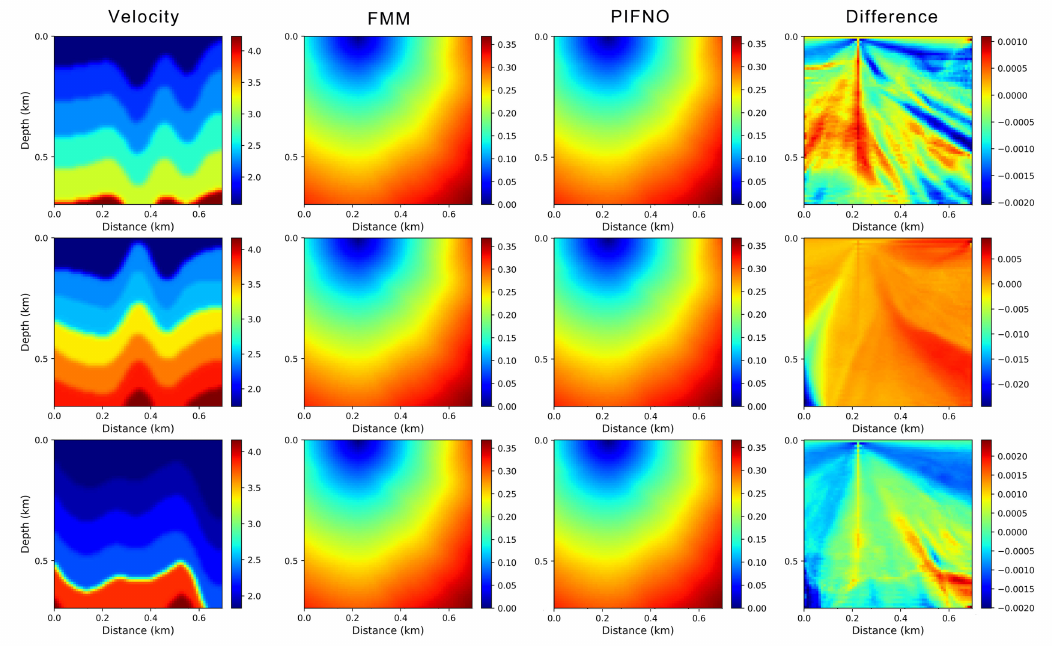} 
\caption{Training data from the CurveVel-A family: velocity models (first column), numerical traveltime from FMM (second column), predicted traveltime from PIFNO (third column), and traveltime difference (fourth column) for a source on the left .}
\label{fig:curvelet50_train_T_isou3}
\end{center}   
\end{figure}

\begin{figure}
\begin{center}
\includegraphics[width=1.0\textwidth]{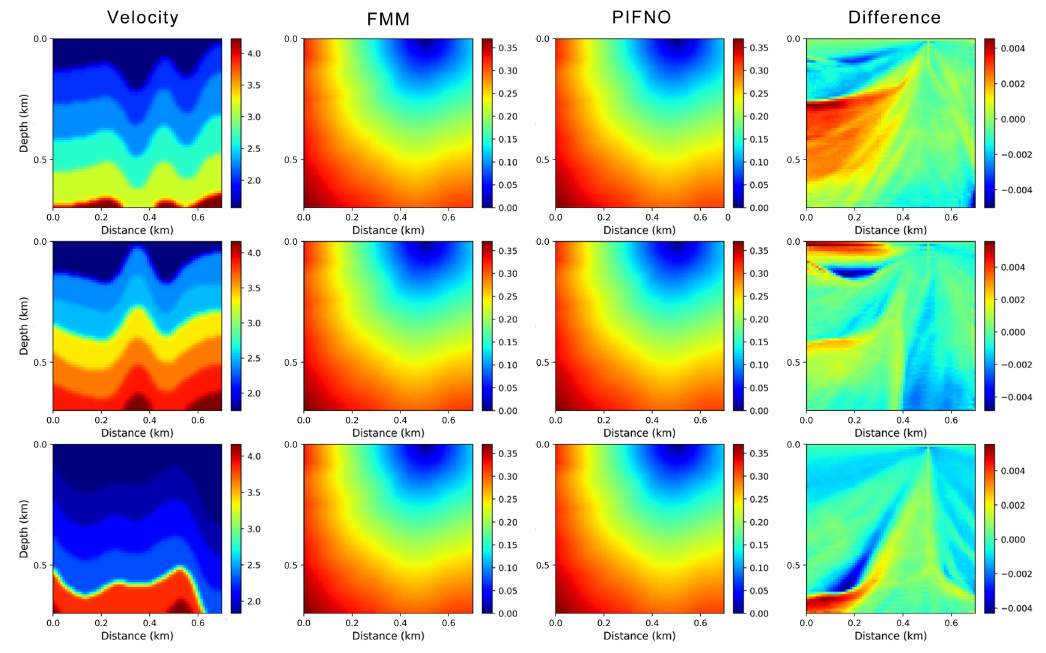} 
\caption{Training data from the CurveVel-A family: velocity models (first column), numerical traveltime from FMM (second column), predicted traveltime from PIFNO (third column), and traveltime difference (fourth column) for a source on the right.}
\label{fig:curvelet50_train_T_isou7}
\end{center}   
\end{figure}

Subsequently, Fig. \ref{fig:curvelet50_train_T} provides a comparison of traveltimes. It illustrates both FMM-calculated and PIFNO-predicted traveltimes, with the source positioned centrally. Figs. \ref{fig:curvelet50_train_T_isou3} and \ref{fig:curvelet50_train_T_isou7} extend this comparison, depicting traveltimes for sources located on the left and right sides, respectively. These figures collectively indicate that PIFNO, once trained, effectively predicts traveltimes across various velocity models and source locations.

\begin{figure}
\begin{center}
\includegraphics[width=1.0\textwidth]{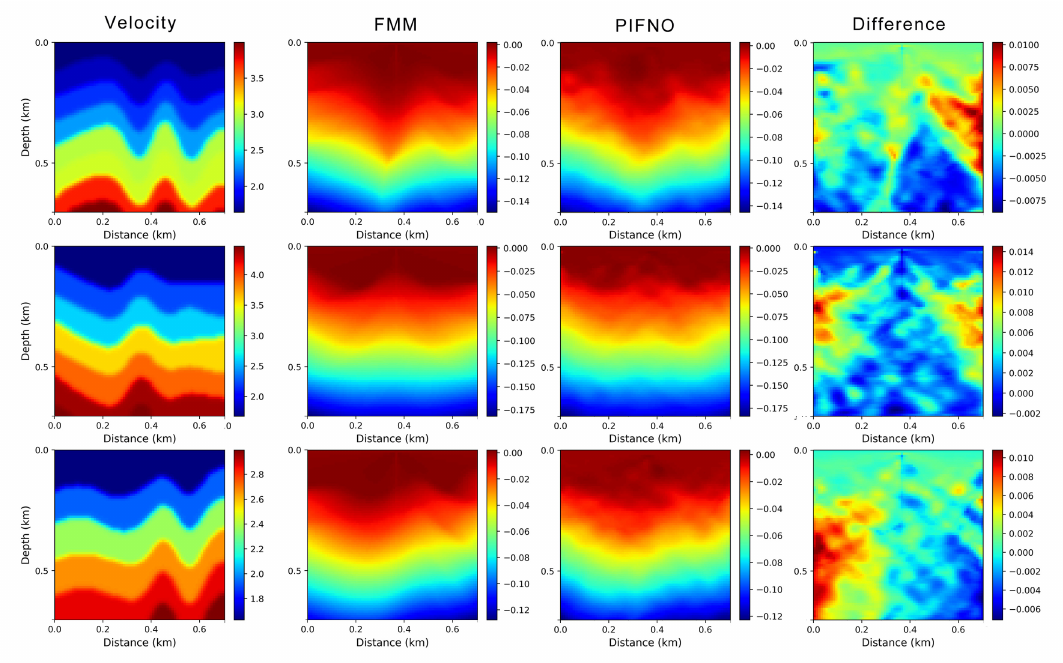} 
\caption{Test data from the CurveVel-A family: velocity models (first column), numerical $\tau$ from FMM (second column), predicted $\tau$ from PIFNO (third column), and $\tau$ difference (fourth column).}
\label{fig:curvelet10_test_tau}
\end{center}   
\end{figure}

\begin{figure}
\begin{center}
\includegraphics[width=1.0\textwidth]{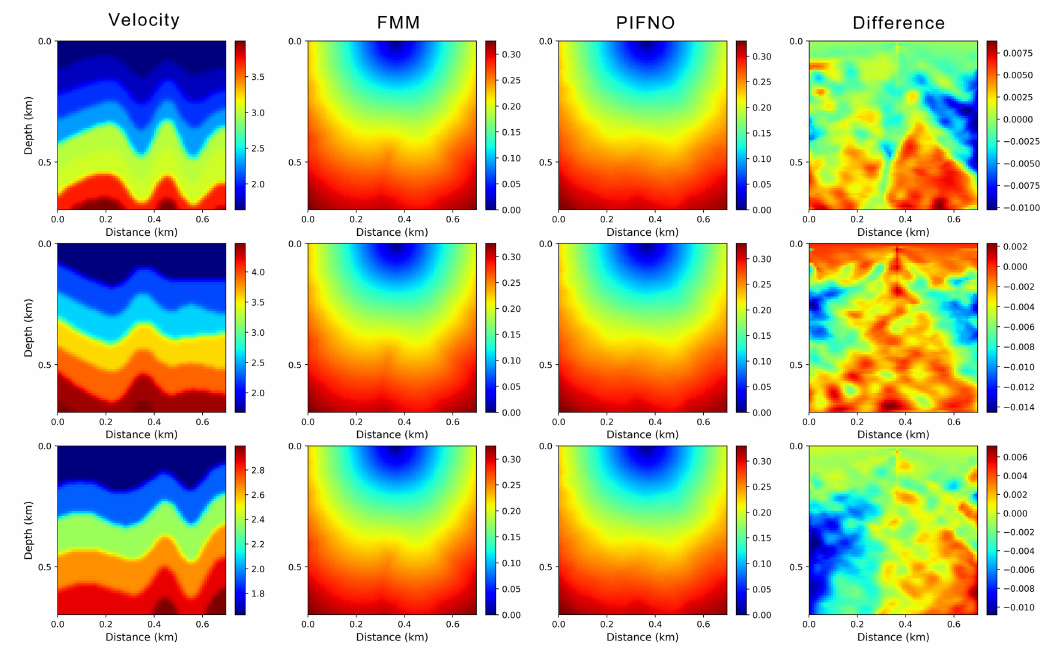} 
\caption{Test data from the CurveVel-A family: velocity models (first column), numerical traveltime from FMM (second column), predicted traveltime from PIFNO (third column), and traveltime difference (fourth column).}
\label{fig:curvelet10_test_T}
\end{center}   
\end{figure}

To further examine PIFNO's generalization capabilities across different velocity models, we employed three additional models from the CurveVel-A family, not included in the training. These models are displayed in the first column of Fig. \ref{fig:curvelet10_test_tau}. Our comparison between FMM-calculated and PIFNO-predicted $\tau$, shown in Fig. \ref{fig:curvelet10_test_tau}'s  second and third columns, reveals an almost identical general distribution, with only minor differences in some areas. Similarly, the comparison of traveltimes $T$, calculated by FMM and predicted by PIFNO, are depicted in the second and third columns of Fig. \ref{fig:curvelet10_test_T}. Here again, differences are mild, as evidenced by the data in Fig. \ref{fig:curvelet10_test_T}'s fourth column.

In our initial experiment, we achieved reasonably good results in predicting traveltimes using test velocity models not included in the training dataset. However, these models belonged to the same velocity family as the training set. To more rigorously evaluate the velocity generalization capabilities of the Physics-Informed Fourier Neural Operator PIFNO, we proceeded to test it using different velocity families sourced from OpenFWI.

\begin{figure}
\begin{center}
\includegraphics[width=1.0\textwidth]{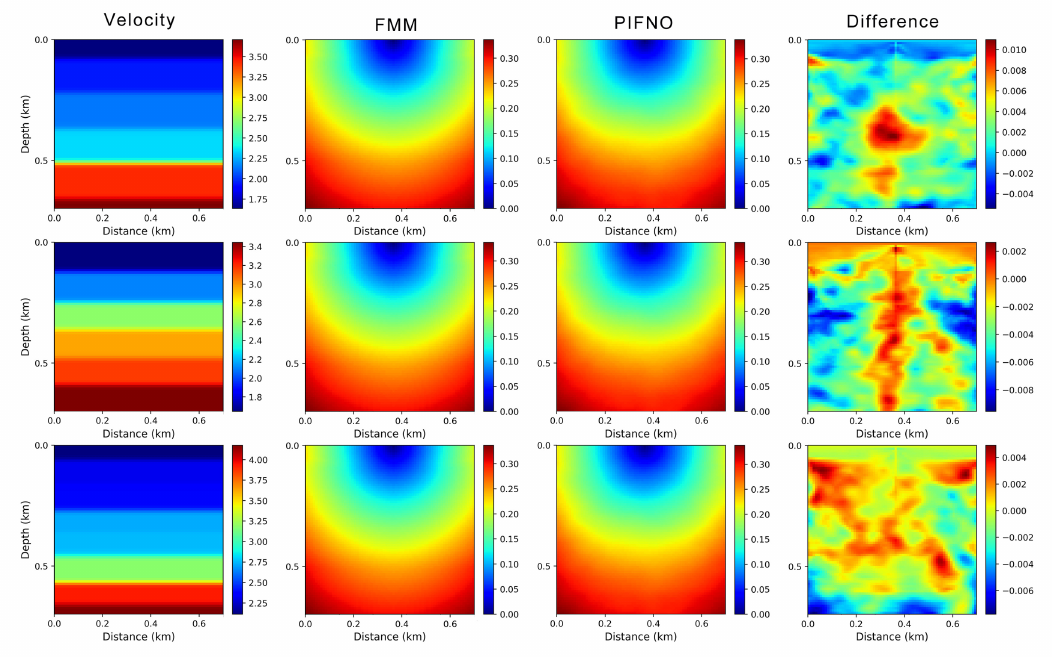} 
\caption{Test data from the layer family: velocity models (first column), numerical traveltime from fast sweeping (second column), predicted traveltime from PIFNO (third column), and traveltime difference for a source on the right (fourth column).}
\label{fig:curvelet_train_layer_test_T}
\end{center}   
\end{figure}

Our first set of tests involved models from the FlatVel-A family. Fig. \ref{fig:curvelet_train_layer_test_T} outlines the same setup: its first column displays three velocity models characterized by flat layers. The second and third columns compare traveltimes calculated using FMM and those predicted by PIFNO. Their near-identical nature underscores PIFNO's accuracy in handling this test data, as corroborated by the minimal discrepancies evident in the fourth column of Fig. \ref{fig:curvelet_train_layer_test_T}.

\begin{figure}
\begin{center}
\includegraphics[width=1.0\textwidth]{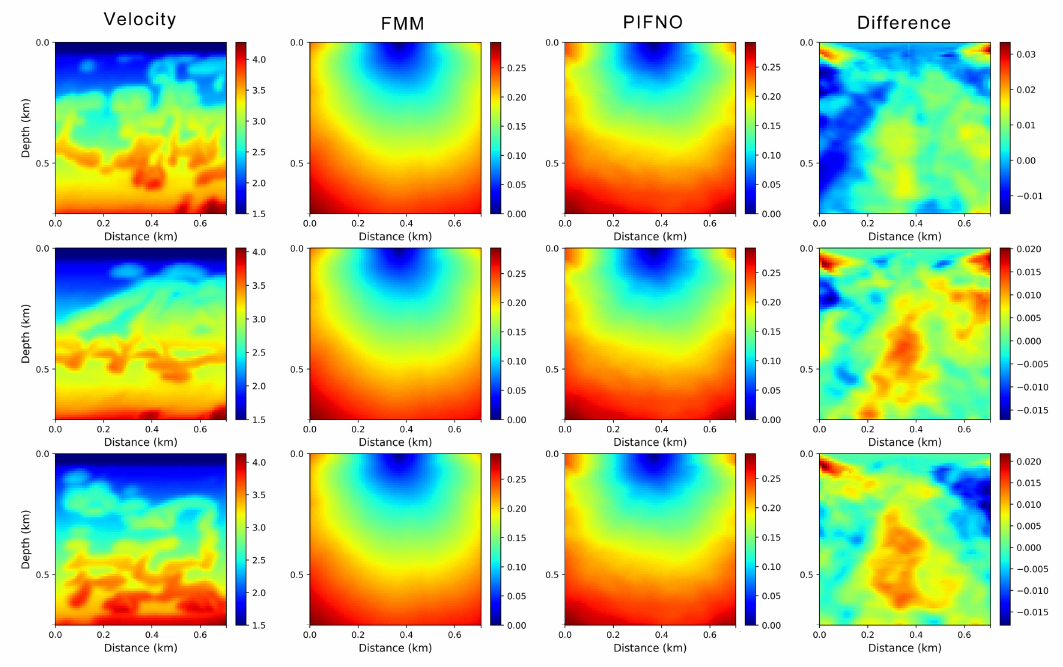} 
\caption{Test data from the Style-A family: velocity models (first column), numerical traveltime from fast sweeping (second column), predicted traveltime from PIFNO (third column), and traveltime difference for a source on the right (fourth column).}
\label{fig:curvelet_train_model6_test_T}
\end{center}   
\end{figure}

Subsequently, we challenge PIFNO by applying it to the Style-A velocity family, a set with significantly divergent structural features compared to those previously tested. The distinct structures of these models are displayed in the first column of Fig. \ref{fig:curvelet_train_model6_test_T}. A comparison of FMM-calculated and PIFNO-predicted traveltimes is provided in the second and third columns of the same figure. Here, we noted a more pronounced difference between the calculated and predicted values, with a slightly larger discrepancy than observed in the earlier tests. This increased difference is attributable to the stark contrast in structural features between the Style-A family and the CurveVel-A family used during training. Consequently, the trained PIFNO exhibited some difficulties in accurately recognizing and adapting to the velocity model characteristics of this new, unfamiliar velocity family.

\section{Discussion}

The PIFNO architecture we developed for solving seismic traveltimes primarily comprises two parts: FNO blocks and a multilayer neural network. The FNO blocks effectively extract features from the input velocity models and background traveltimes. According to the universal approximation theorem  \cite{hornik1991approximation}, the multilayer neural network component is tasked with approximating the target output $\tau$. The integration of physics-informed approaches into our FNO architecture circumvents the extensive requirement for labeled training data typically necessary for neural operators, as it allows the system to leverage underlying physical principles to guide the learning process, thus enabling the neural network to approximate the target output $\tau$ with greater data efficiency.

Differing from PINNs that employ automatic differentiation to compute function derivatives, PIFNO leverages a finite-difference scheme for the same purpose. This scheme not only provides satisfactory accuracy but also enhances the training process by accelerating convergence rates. Unlike automatic differentiation, which operates on random individual points and may struggle to identify the optimal direction for network optimization, finite difference benefits from linking and utilizing adjacent point information, thereby simplifying the optimization process \cite{fang2021high}.

PIFNO's input comprises both the velocity models and background traveltimes, with the latter encapsulating information about the source locations. PIFNO's training regime is entirely unsupervised, relying solely on the principles of physics. The PIFNO-based factorized eikonal solver we propose is capable of generating multi-source traveltimes for not only the velocity models used in training but can also adapt to test velocity models exhibiting similar structural characteristics. Enhancing the model's ability to generalize across various velocities could be achieved by incorporating a broader range of velocity models with diverse structural features into the training set. By assigning velocity information to the batch dimension, the architecture allows for the inclusion of as many velocity models as a mini-batch training strategy can be applied.
 
\section{Conclusion}

We have developed an innovative multi-source seismic traveltime simulation method adaptable to various velocity models, employing an advanced deep-learning technique known as the physics-informed Fourier neural operator (PIFNO). We proposed to solve a factorized eikonal equation that separates the seismic traveltime to a background term and a perturbation term. In preparing the input tensor, we integrate the velocity model with the background traveltime, where the latter reflects the positions of seismic sources. The factorized eikonal equation serves as the foundation for our loss function in the optimization of PIFNO. The trained PIFNO establishes a non-linear relationship between a composite of the velocity and background traveltime, and the perturbation traveltime. This capability enables PIFNO to adeptly predict seismic traveltimes for a spectrum of velocities across multiple sources.

\section{Acknowledgement}

We thank Jilin University for its support. 

\section{Data availability}

We will share the codes related upon acceptance of this paper.

\bibliographystyle{unsrt}  
\bibliography{refs/pifno}

\begin{thebibliography}{10}

\bibitem{Yilmaz2001Seismic}
Özdoğan Yilmaz and Stephen~M Doherty.
\newblock {\em Seismic data analysis : processing, inversion, and
  interpretation of seismic data /-(2nd ed.)}.
\newblock Seismic data analysis : processing, inversion, and interpretation of
  seismic data /-(2nd ed.), 2001.

\bibitem{li2013first}
Siwei Li, Alexander Vladimirsky, and Sergey Fomel.
\newblock First-break traveltime tomography with the double-square-root eikonal
  equation.
\newblock {\em Geophysics}, 78(6):U89--U101, 2013.

\bibitem{izzatullah2022laplace}
Muhammad Izzatullah, Isa~Eren Yildirim, Umair~Bin Waheed, and Tariq Alkhalifah.
\newblock Laplace hypopinn: physics-informed neural network for hypocenter
  localization and its predictive uncertainty.
\newblock {\em Machine Learning: Science and Technology}, 3(4):045001, 2022.

\bibitem{sethian1996fast}
James~A Sethian.
\newblock A fast marching level set method for monotonically advancing fronts.
\newblock {\em proceedings of the National Academy of Sciences},
  93(4):1591--1595, 1996.

\bibitem{zhao2005fast}
Hongkai Zhao.
\newblock A fast sweeping method for eikonal equations.
\newblock {\em Mathematics of computation}, 74(250):603--627, 2005.

\bibitem{mousavi2022deep}
S~Mostafa Mousavi and Gregory~C Beroza.
\newblock Deep-learning seismology.
\newblock {\em Science}, 377(6607):eabm4470, 2022.

\bibitem{wu2023sensing}
Xinming Wu, Jianwei Ma, Xu~Si, Zhengfa Bi, Jiarun Yang, Hui Gao, Dongzi Xie,
  Zhixiang Guo, and Jie Zhang.
\newblock Sensing prior constraints in deep neural networks for solving
  exploration geophysical problems.
\newblock {\em Proceedings of the National Academy of Sciences},
  120(23):e2219573120, 2023.

\bibitem{raissi2019physics}
Maziar Raissi, Paris Perdikaris, and George~E Karniadakis.
\newblock Physics-informed neural networks: A deep learning framework for
  solving forward and inverse problems involving nonlinear partial differential
  equations.
\newblock {\em Journal of Computational physics}, 378:686--707, 2019.

\bibitem{kovachki2021neural}
Nikola Kovachki, Zongyi Li, Burigede Liu, Kamyar Azizzadenesheli, Kaushik
  Bhattacharya, Andrew Stuart, and Anima Anandkumar.
\newblock Neural operator: Learning maps between function spaces.
\newblock {\em arXiv preprint arXiv:2108.08481}, 2021.

\bibitem{alkhalifah2021wavefield}
Tariq Alkhalifah, Chao Song, Umair bin Waheed, and Qi~Hao.
\newblock Wavefield solutions from machine learned functions constrained by the
  helmholtz equation.
\newblock {\em Artificial Intelligence in Geosciences}, 2:11--19, 2021.

\bibitem{song2021solving}
Chao Song, Tariq Alkhalifah, and Umair~Bin Waheed.
\newblock Solving the frequency-domain acoustic vti wave equation using
  physics-informed neural networks.
\newblock {\em Geophysical Journal International}, 225(2):846--859, 2021.

\bibitem{bin2021pinneik}
Umair bin Waheed, Ehsan Haghighat, Tariq Alkhalifah, Chao Song, and Qi~Hao.
\newblock Pinneik: Eikonal solution using physics-informed neural networks.
\newblock {\em Computers \& Geosciences}, 155:104833, 2021.

\bibitem{waheed2021pinntomo}
Umair~Bin Waheed, Tariq Alkhalifah, Ehsan Haghighat, Chao Song, and Jean
  Virieux.
\newblock Pinntomo: Seismic tomography using physics-informed neural networks.
\newblock {\em arXiv preprint arXiv:2104.01588}, 2021.

\bibitem{rasht2022physics}
Majid Rasht-Behesht, Christian Huber, Khemraj Shukla, and George~Em
  Karniadakis.
\newblock Physics-informed neural networks (pinns) for wave propagation and
  full waveform inversions.
\newblock {\em Journal of Geophysical Research: Solid Earth},
  127(5):e2021JB023120, 2022.

\bibitem{song2022wavefield}
Chao Song and Tariq~A Alkhalifah.
\newblock Wavefield reconstruction inversion via physics-informed neural
  networks.
\newblock {\em IEEE Transactions on Geoscience and Remote Sensing}, 60:1--12,
  2022.

\bibitem{deng2021openfwi}
Chengyuan Deng, Shihang Feng, Hanchen Wang, Xitong Zhang, Peng Jin, Yinan Feng,
  Qili Zeng, Yinpeng Chen, and Youzuo Lin.
\newblock {OpenFWI:} large-scale multi-structural benchmark datasets for
  seismic full waveform inversion.
\newblock {\em arXiv preprint arXiv:2111.02926}, 2021.

\bibitem{julian1977three}
BR~Julian, Db~Gubbins, et~al.
\newblock Three-dimensional seismic ray tracing.
\newblock {\em Journal of Geophysics}, 43(1):95--113, 1977.

\bibitem{cerveny2001seismic}
Vlastislav Cerven{\`y}.
\newblock {\em Seismic ray theory}, volume 110.
\newblock Cambridge university press Cambridge, 2001.

\bibitem{li2020fourier}
Zongyi Li, Nikola Kovachki, Kamyar Azizzadenesheli, Burigede Liu, Kaushik
  Bhattacharya, Andrew Stuart, and Anima Anandkumar.
\newblock Fourier neural operator for parametric partial differential
  equations.
\newblock {\em arXiv preprint arXiv:2010.08895}, 2020.

\bibitem{yang2021seismic}
Yan Yang, Angela~F Gao, Jorge~C Castellanos, Zachary~E Ross, Kamyar
  Azizzadenesheli, and Robert~W Clayton.
\newblock Seismic wave propagation and inversion with neural operators.
\newblock {\em The Seismic Record}, 1(3):126--134, 2021.

\bibitem{song2022high}
Chao Song and Yanghua Wang.
\newblock High-frequency wavefield extrapolation using the fourier neural
  operator.
\newblock {\em Journal of Geophysics and Engineering}, 19(2):269--282, 2022.

\bibitem{wei2022small}
Wei Wei and Li-Yun Fu.
\newblock Small-data-driven fast seismic simulations for complex media using
  physics-informed fourier neural operators.
\newblock {\em Geophysics}, 87(6):T435--T446, 2022.

\bibitem{skfmm}
J.~Furtney et~al.
\newblock scikit-fmm: the fast marching method for python.
\newblock {\em https://github.com/scikit-fmm/scikit-fmm,}, 2015.

\bibitem{hornik1991approximation}
Kurt Hornik.
\newblock Approximation capabilities of multilayer feedforward networks.
\newblock {\em Neural networks}, 4(2):251--257, 1991.

\bibitem{fang2021high}
Zhiwei Fang.
\newblock A high-efficient hybrid physics-informed neural networks based on
  convolutional neural network.
\newblock {\em IEEE Transactions on Neural Networks and Learning Systems},
  33(10):5514--5526, 2021.

\end{thebibliography}

\end{document}